\documentclass[manuscript]{aastex}

\usepackage{color,ulem}

\newcommand{\simlt}{\lower.5ex\hbox{$\; \buildrel < \over \sim \;$}}

\shortauthors{Ishikawa et al.}

\begin{document}

\title{On the inversion of the scattering polarization
and the Hanle effect signals in the hydrogen Lyman-$\alpha$ line}

\author{R. Ishikawa\altaffilmark{1}, A. Asensio Ramos\altaffilmark{2},
L. Belluzzi\altaffilmark{3,4}, R. Manso Sainz\altaffilmark{2}, J. {\v S}t{\v e}p{\'a}n\altaffilmark{5},
J. Trujillo Bueno\altaffilmark{2}, M. Goto\altaffilmark{6},
and S. Tsuneta\altaffilmark{7}}
\altaffiltext{1}{National Astronomical
Observatory of Japan, 2-21-1 Osawa, Mitaka, Tokyo 181-8588, Japan}
\altaffiltext{2}{Instituto de Astrofisica de Canarias, 38205 La Laguna, Tenerife, Spain}
\altaffiltext{3}{Istituto Ricerche Solari Locarno (IRSOL), 6605 Locarno Monti, Switzerland}
\altaffiltext{4}{Kiepenheuer-Institut f\"ur Sonnenphysik, Freiburg, Germany}
\altaffiltext{5}{Astronomical Institute of the Academy of Sciences, Fri\v{c}ova
298, 25165 Ond\v{r}ejov, Czech Republic}
\altaffiltext{6}{National Institute for Fusion Science, Toki 509-5292, Japan}
\altaffiltext{7}{Institute of Space and Astronautical Science, Japan Aerospace Exploration Agency, 3-1-1 Yoshinodai, Chuo, Sagamihara, Kanagawa 252-5210, Japan}
\email{ryoko.ishikawa@nao.ac.jp}

\begin{abstract}
Magnetic field measurements in the upper chromosphere and above, where the gas-to-magnetic pressure ratio $\beta$ is lower than unity, are essential for understanding the thermal structure and dynamical activity of the solar atmosphere. Recent developments in the theory and numerical modeling of polarization in spectral lines have suggested that 
information on the magnetic field of the chromosphere-corona transition region could be obtained by measuring the linear polarization of the solar disk radiation at the core of the hydrogen Lyman-$\alpha$ line at 121.6~nm, which is produced by scattering processes and the Hanle effect. The Chromospheric Lyman-$\alpha$ Spectropolarimeter (CLASP) sounding rocket experiment aims to measure the intensity (Stokes $I$) and the linear polarization profiles ($Q/I$ and $U/I$) of the hydrogen Lyman-$\alpha$ line. In this paper we clarify the information that the Hanle effect can provide by
applying a Stokes inversion technique based on a database search. The database contains all theoretical $Q/I$ and $U/I$ profiles calculated in a one-dimensional semi-empirical model of the solar atmosphere for all possible values of the strength, inclination, and azimuth of the magnetic field vector, though this atmospheric region is highly inhomogeneous and dynamic. We focus on understanding the sensitivity of the inversion results to the noise and spectral resolution of the synthetic observations as well as the ambiguities and limitation inherent to the Hanle effect when only the hydrogen Lyman-$\alpha$ is used. 
We conclude that spectropolarimetric observations with CLASP can indeed be a suitable diagnostic tool for probing the magnetism of the transition region, especially when complemented with information on the magnetic field azimuth that can be obtained from other instruments.
\end{abstract}


\section{Introduction}
The chromosphere and the transition region of the Sun lie 
between the cooler photosphere, where the ratio of gas to magnetic pressure
$\beta>1$, and the $10^6$ K corona, where $\beta<1$.
It is believed that in this interface region the magnetic forces start to dominate over the hydrodynamic forces, and that local energy dissipation and energy transport 
to the upper layers via various fundamental plasma processes are taking place.
Recent observations \citep[e.g.,][]{Shibata2007,Katsukawa2007,DePontieu2007,Okamoto2007,Okamoto2011,Vecchio2009}
have revealed ubiquitous dynamical chromospheric
activities such as jets, Alfv\'enic waves, and shocks,
which are thought to play a key
role in the heating of the chromosphere and corona and in
the acceleration of the solar wind.
However, we do not have any significant empirical knowledge on
the strength and direction of the magnetic field
in the upper solar chromosphere and transition region.

The information on the magnetic field of the solar atmosphere
is encoded in the polarization that some physical mechanisms introduce in the
spectral lines. The familiar Zeeman effect can introduce 
polarization in the spectral lines that originate in the upper solar chromosphere and the transition region.
However, because such lines are broad and the magnetic field there is expected to be rather  weak, the induced polarization amplitudes will be very small (except perhaps in sunspots), and the Zeeman effect has limited applicability. Fortunately, the Hanle effect \citep[the magnetic-field-induced modification of the linear polarization caused by scattering processes in a spectral line,][]{Casini2008} in some of the allowed UV lines that originate in the upper chromosphere and transition region is expected to be a more suitable diagnostic tool \citep{Trujillo2011,Trujillo2012,Belluzzi2012}.

The hydrogen Lyman-$\alpha$ line ($\lambda=121.567$~nm)
is particularly suitable because (1) the line-core polarization originates at the base of the solar transition region, where ${\beta}{\ll}1$
\citep{Trujillo2011,Belluzzi2012b,Stepan2012}, (2) collisional depolarization plays a rather 
insignificant role \citep[e.g.,][]{Stepan2011}, and (3) via the Hanle effect the
scattering polarization is sensitive to the magnetic fields
expected for the upper chromosphere and transition region \citep{Trujillo2011}.

The Chromospheric Lyman-Alpha Spectropolarimeter (CLASP) is a sounding rocket experiment developed by researchers from Japan, USA, and Europe
\citep[][]{Ishikawa2011,Narukage2011,Kano2012,Kobayashi2012},
which is expected to fly in 2015.
The first, very important goal of this sounding rocket experiment is the  
measurement of the linear polarization signals
produced by scattering processes in the Lyman-$\alpha$ line.
The second goal is the detection of the Hanle effect action on the core of $Q/I$ and $U/I$ in order to 
constrain the magnetic field of the transition region from the observed Stokes profiles.
CLASP will measure the linear polarization profiles of the Lyman-$\alpha$ line
within a spectral window of at least
$\pm0.05$~nm around the line center, where in addition to the line core polarization itself (where the Hanle effect operates),
we expect the largest linear polarization signals produced by
the joint action of partial frequency redistribution and $J-$state
interference effects \citep{Belluzzi2012b}.
Polarization sensitivities
of 0.1\% and 0.5\% are required in the line core (i.e., $\pm0.02$~nm
around the line center) and in the line wings (at $>\pm0.05$~nm), respectively.
In order to achieve these polarization sensitivities,
the $400\arcsec$ spectrograph slit will be
fixed at the selected observing target
during the CLASP observation time of $<5$~min. Furthermore,
after the data recovery, we will add consecutive measurements and
perform spatial averaging.

In this paper, we clarify the information we expect to determine with the CLASP experiment, 
providing a strategy suitable for highlighting the
ambiguities of the Hanle effect and the complexity of the ensuing inference problem. 
To this end, we have used a plane-parallel (one-dimensional) semi-empirical model of the solar atmosphere,
and we have created a database of theoretical Stokes $Q/I$ and $U/I$ profiles for all possible strength, inclination, and azimuth values of the magnetic field vector.
Then, we investigate the possibility of recovering the magnetic field information using the characteristics of CLASP (noise level, spectral resolution, etc.)
and the ambiguities and limitation inherent to the Hanle effect. 
We also discuss the most suitable observing targets and data analysis strategy for constraining the magnetic field information.

The ambiguities inherent to magnetic field diagnostics can be reduced by exploiting
the joint action of the Hanle and Zeeman effect, where both linear and circular
polarization signals are used to constrain the magnetic field vector
\citep[][see also \citealt{Asensio2008,Centeno2010, Anusha2011}]{Landi1993,Landi1982,Trujillo2002,Lopez2003}.
Unfortunately, while the Lyman-$\alpha$ line is most advantageous to explore
the magnetism of the various regions in the transition region (where $\beta\ll1$),
it is challenging to measure the contribution of the Zeeman effect to the circular polarization in UV lines.
Here we propose alternative ways to alleviate this issue in the subsequent sections.

We assume that the quiet solar atmosphere can be represented by   
a plane-parallel semi-empirical model atmosphere, 
even though the upper solar chromosphere is a highly inhomogeneous and dynamic physical system, much more complex than the idealization of the one-dimensional static semi-empirical model used here. Such inhomogeneity and dynamics causes larger amplitudes and spatial variations in the scattering polarization signals \citep{Stepan2012,Stepan2014}.
However, it is of interest to note that
\citet{Trujillo2011} showed that the amplitude and shape of the $Q/I$ profiles calculated in the  quiet-Sun plane-parallel semi-empirical model of \citet{Fontenla1993}
are qualitatively similar to the temporally-averaged profiles obtained from the Stokes $I$ and $Q$ signals computed at each time step of the chromospheric hydro-dynamical model of \citet{Carlsson1997}. Moreover, spatial and temporal averaging of the scattering polarization calculated in three-dimensional (3D) atmospheric models tend to produce $Q/I$ Lyman-$\alpha$ 
signals more or less similar to those calculated in plane-parallel semi-empirical model atmospheres \citep{Stepan2014}. 
In the case of the CLASP experiment, which will require both spatial ($\sim10\arcsec$) and temporal averaging ($\sim$5~min) to attain the necessary signal-to-noise ratio,
the model atmosphere for a first approximate interpretation of  
the CLASP data could be a plane-parallel semi-empirical model.
We believe that the post-launch data analysis will pave the way for further improvements, for example via forward modeling calculations of the Lyman-$\alpha$
scattering polarization signals using increasingly realistic 3D models of the solar chromosphere.

\section{Scattering polarization and the Hanle effect}

The expected scattering polarization in the core of the 
hydrogen Lyman-$\alpha$ line physically originates from population imbalances and 
quantum coherence between the magnetic sublevels pertaining to the $2p$ 
${^2{\rm P}}_{3/2}$ upper level, both of which are produced from the absorption of anisotropic 
radiation by the hydrogen atoms of the solar transition region.

Typically, in weakly magnetized stellar atmospheres, the absorption of 
anisotropic radiation produces atomic level alignment (i.e., population 
imbalances between magnetic sublevels having different $|M|$ values, $M$ being 
the magnetic quantum number). On the other hand, atomic level orientation (i.e., population 
imbalances between sublevels with $M > 0$ and $M < 0$) can be neglected 
in the modeling of the Lyman-$\alpha$ linear polarization. 
The anisotropy and symmetry properties of the radiation field that illuminates 
the atomic system can be conveniently quantified through the so-called 
radiation field tensor $J^K_Q$ \citep[see Eq. (5.157) of ][]{Landi2004}. 
If the radiation field has cylindrical symmetry along a given 
direction, and it has no circular polarization, then, if the quantization axis 
is taken along the symmetry axis, only the components $J^0_0$ and $J^2_0$ are 
non-zero. The former represents the mean intensity of the radiation field, 
while the latter quantifies its degree of anisotropy. The explicit expression
of the $J^2_0$ component of the frequency-integrated radiation field tensor is 
given by 

\begin{equation}
	J^2_0 = \int {\rm d}x \oint \frac{{\rm d} \Omega}{4 \pi} 
	\frac{\phi_x}{2 \sqrt{2}} \left[ (3 \mu^2 - 1) I_{x \Omega} + 
	3(1 - \mu^2) Q_{x \Omega} \right] \; ,
\end{equation}
where $\phi_x$ is the absorption profile, $x$ is the normalized frequency distance 
from line center, and $\mu = \cos\theta$ (with $\theta$ the angle between the 
radiation beam under consideration and the quantization axis).
In the solar atmosphere, the contribution of the Stokes $Q_{x \Omega}$ 
parameter to $J^2_0$ is very small compared with that caused by the specific 
intensity $I_{x \Omega}$. 
Thus, in analogy with the spherical harmonics $Y^0_2$, the $J^2_0$ tensor 
represents whether the local illumination of the atomic system is dominated by 
predominantly vertical ($J^2_0 > 0$) or predominantly horizontal ($J^2_0 < 0$) 
illumination. 
In the case of $J^2_0 > 0$ ($J^2_0 < 0$), photon absorption processes take place 
predominantly through $\Delta M = \pm 1$ ($\Delta M = 0$) transitions, which gives rise to 
population imbalances among the various magnetic sublevels.

In this work, we consider a plane-parallel atmosphere, whose parameters depend 
only on the height. 
Taking the quantization axis along the vertical, it can be shown that in 
the absence of magnetic fields, or in the presence of a vertical magnetic 
field, any coherence between pairs of magnetic sublevels of the $2p$ 
${^2{\rm P}}_{3/2}$ upper level is zero, while it is non-zero if an 
inclined magnetic field is present. 

Although choosing the quantization axis along the local vertical can be 
advantageous in some cases, in order to investigate the impact of the magnetic 
field on the atomic polarization (i.e., the Hanle effect), it is convenient to 
choose it along the magnetic field direction.
In this case, describing atomic polarization through the multipole moments of 
the density matrix, $\rho^K_Q$ \citep[e.g., Sect.~3.7 of][]{Landi2004}, the effect of the magnetic field is described by the equation \citep[see Sect.~10.3 of][]{Landi2004}
\begin{equation}
	\rho^K_Q(J_u) = \frac{1}{1 + {\rm i} Q \Gamma_u}  
	\left[ \rho^K_Q(J_u) \right]_{B=0} \; ,
\end{equation}
where $\Gamma_u = 8.79 \times 10^6 \, B \, g_{J_u} / A_{u \ell}$ ($B$ is the 
field strength in gauss, $g_{J_u}$ is the Land\'e factor, and $A_{u \ell}$ is 
the Einstein coefficient for the spontaneous emission process in s$^{-1}$), and
$[\rho^K_Q(J_u)]_{B=0}$ represent the value of the $\rho^K_Q(J_u)$ elements for 
the non-magnetic case. 
This equation shows that in the magnetic field reference frame the population 
imbalances represented by the $\rho^K_Q(J_u)$ elements with $Q = 0$ are 
unaffected by the magnetic field, while the elements with $Q \ne 0$, indicating 
the atomic coherence, are reduced and dephased with respect to the non-magnetic 
case. 
The Hanle effect can be thus be defined as the modification of the atomic-level 
polarization (in particular the modification of coherence) and the ensuing
observables effects on the emergent Stokes $Q$ and $U$ profiles, caused by the 
action of an inclined magnetic field. 
For the hydrogen Lyman-$\alpha$ line, the magnetic field strength for which 
$\Gamma_u = 1$ (i.e., the critical magnetic field for the onset of the Hanle 
effect) is $B = 53$~G.

\section{Database}
\label{database}
The Lyman-$\alpha$ line consists of two blended transitions between
the $1p$~$^2{\rm S}_{1/2}$ lower level
and the $2p$~$^2{\rm P}_{1/2}$ and $2p$~$^2{\rm P}_{3/2}$ upper levels.
To estimate the linear polarization
of the Lyman-$\alpha$ line,
we follow the approach of \citet{Trujillo2011}, and
we provide a quick overview here (refer to \citet{Trujillo2011} for details).
We consider the quiet-Sun semi-empirical model of \citet[][]{Fontenla1993}, which is hereafter referred to as the FAL-C model.
Thus, our model atmosphere is plane-parallel, and the physical quantities only depend on the coordinate $Z$. 
The hydrogen atomic model we have used includes the fine structure of
the first two $n-$levels of the hydrogen, where $n$ is the
principal quantum number.
The excitation state of
each level is quantified by means of the
multipolar components of the atomic density matrix,
which are self-consistently obtained by solving
the statistical equilibrium equations
and the Stokes vector radiative transfer equation
\citep[see chapter 7 of ][]{Landi2004}.
Thus, we assume complete frequency re-distribution (CRD),
which is a suitable approximation for the estimation
of the scattering polarization at the Lyman-$\alpha$ line center \citep{Belluzzi2012b}. 
Isotropic collisions with protons and electrons are also taken into account,
but these collisions have a negligible depolarizing effect on the scattering
polarization of the hydrogen Lyman-$\alpha$ line at the low plasma densities
of the upper chromosphere and the transition region, 
as discussed by \citet{Stepan2011}.

\begin{figure}
\epsscale{0.9}
\plotone{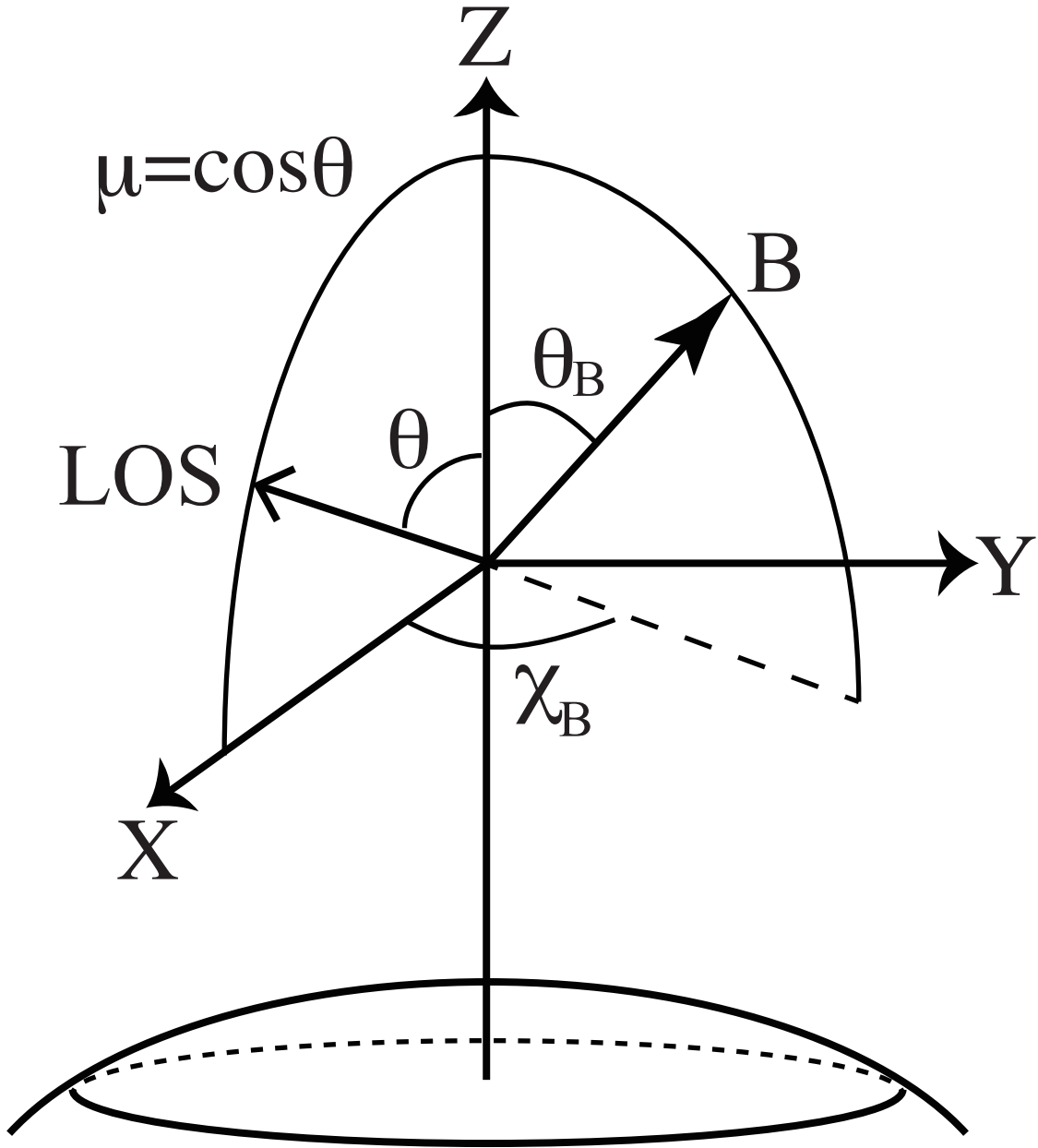}
\caption{Geometry for the scattering event. The $Z$-axis is
normal to the solar surface.
The magnetic field vector is characterized by its magnitude $B$,
inclination angle $\theta_B$, and azimuth $\chi_B$.
The LOS lies in the $X-Z$ plane.
We choose the $Y$-axis as positive reference direction for
Stokes $Q$, which is parallel to the nearest solar limb.}
\label{fig:geometry}
\end{figure}

\begin{figure}
\epsscale{1.0}
\plotone{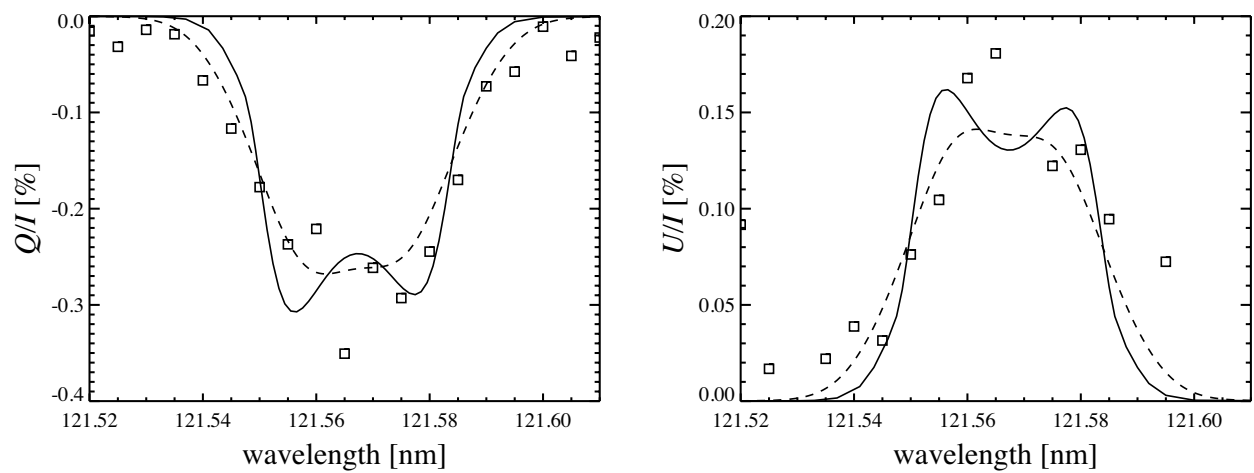}
\caption{The original synthetic Lyman-$\alpha$ $Q/I$ and $U/I$ profiles (solid lines)
for a LOS with $\mu=0.3$, considering the Hanle effect of
a 50~G horizontal magnetic field with an azimuth of $120^{\circ}$.
The $Q/I$ and $U/I$ convolved with a 0.013~nm FWHM Gaussian are shown with dashed lines.
The simulation of the observed profiles (squares),
taking into account 0.005-nm pixel samplings and the random noise of
$\sigma=0.033\%$.}
\label{fig:profile}
\end{figure}

By solving the above-mentioned non-local thermodynamic equilibrium (non-LTE)
radiative transfer problem, we created a database of synthetic Stokes profiles
($I(\lambda)$, $Q(\lambda)$, and $U(\lambda)$)
of the hydrogen Lyman-$\alpha$ line.
We consider the presence of a deterministic magnetic field
of arbitrary strength $B$, inclination $\theta_B$, and azimuth $\chi_B$,
all of which are assumed to be constant with height (Fig.\ref{fig:geometry}).
In total, there are 137751 sets of $Q(\lambda)/I(\lambda)$ and $U(\lambda)/I(\lambda)$
profiles for different magnetic field strengths
($0\le B\le 250~\mathrm{G}$ in $5~\mathrm{G}$ increments),
inclinations
($0^{\circ}\le \theta_B \le 180^{\circ}$ in $5^{\circ}$ increments),
and azimuths
($0^{\circ}\le \chi_B \le 360^{\circ}$ in $5^{\circ}$ increments).
An example of a synthetic profile is given
in Figure~\ref{fig:profile} (solid lines).
For all magnetic parameters, the Stokes $I(\lambda)$ profiles are virtually identical because
only the polarization signals have a measurable sensitivity to the magnetic field.
Here, we consider two scattering geometries: disk center ($\mu=1.0$)
and close-to-the-limb ($\mu=0.3$).

Here, we also simulate the CLASP observations.
The wavelength resolution of the CLASP optics
is $\sim0.013$~nm, and the Stokes spectra are recorded
with a wavelength sampling of 0.005~nm~pixel$^{-1}$.
For a spatial area of less than 10\arcsec,
the polarization sensitivity
is 0.1\% with respect to the intensity at each wavelength pixel
in the line core ($121.567\pm0.02$~nm).
First, the synthetic profiles $I(\lambda),
Q(\lambda)$, and $U(\lambda)$
in the database are
convolved with a 0.013~nm FWHM Gaussian
and are then sampled with a wavelength step of 0.005~nm.
Assuming that the polarization sensitivity will be
achieved at a $3\sigma$ level,
random noise with a standard deviation of $\sigma=0.033\%$
with respect to the intensity at each wavelength bin
is added to the convolved $I$, $Q$, and $U$ profiles.
Finally, we derive the simulated $Q^{obs}(\lambda)/I^{obs}(\lambda)$
and $U^{obs}(\lambda)/I^{obs}(\lambda)$ profiles
(squares in Fig.\ref{fig:profile}).

\section{Hanle diagrams}
\label{hanle_diagram}
\begin{figure}
\epsscale{1.0}
\plotone{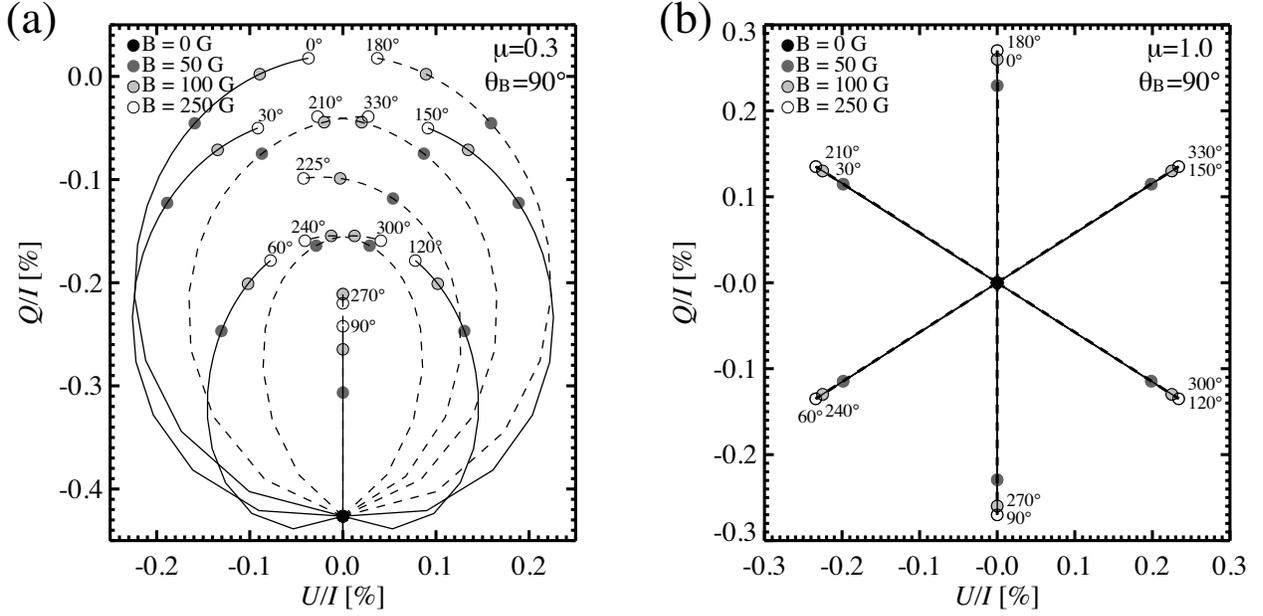}
\caption{The Hanle diagrams of the hydrogen Lyman-$\alpha$ line at a
close-to-the-limb geometry ($\mu=0.3$, (a)) and at the
disk center ($\mu=1.0$, (b)) for the case of
horizontal magnetic fields ($\theta_B=90^{\circ}$).
The curves result from the computation of synthetic
$Q/I$ and $U/I$ line-center signals
for various azimuth $\chi_B$ values and field strengths $B$.
The lines show the variation of $Q/I$ and $U/I$ as a function
of $B$ from $0$~G to $250$~G with constant $\chi_B$.
The black, dark gray, light gray, and white circles
refer to $Q/I$ and $U/I$ for $B=0$, $50$~G, $100$~G, and $250$~G,
respectively.
The values close to the white circle show $\chi_B$ in degrees
for each curve.
The solid and dashed curves represent cases for $0-150^{\circ}$
and $180-330^{\circ}$, respectively.}
\label{fig:diagram1}
\end{figure}

\begin{figure}
\epsscale{1.0}
\plotone{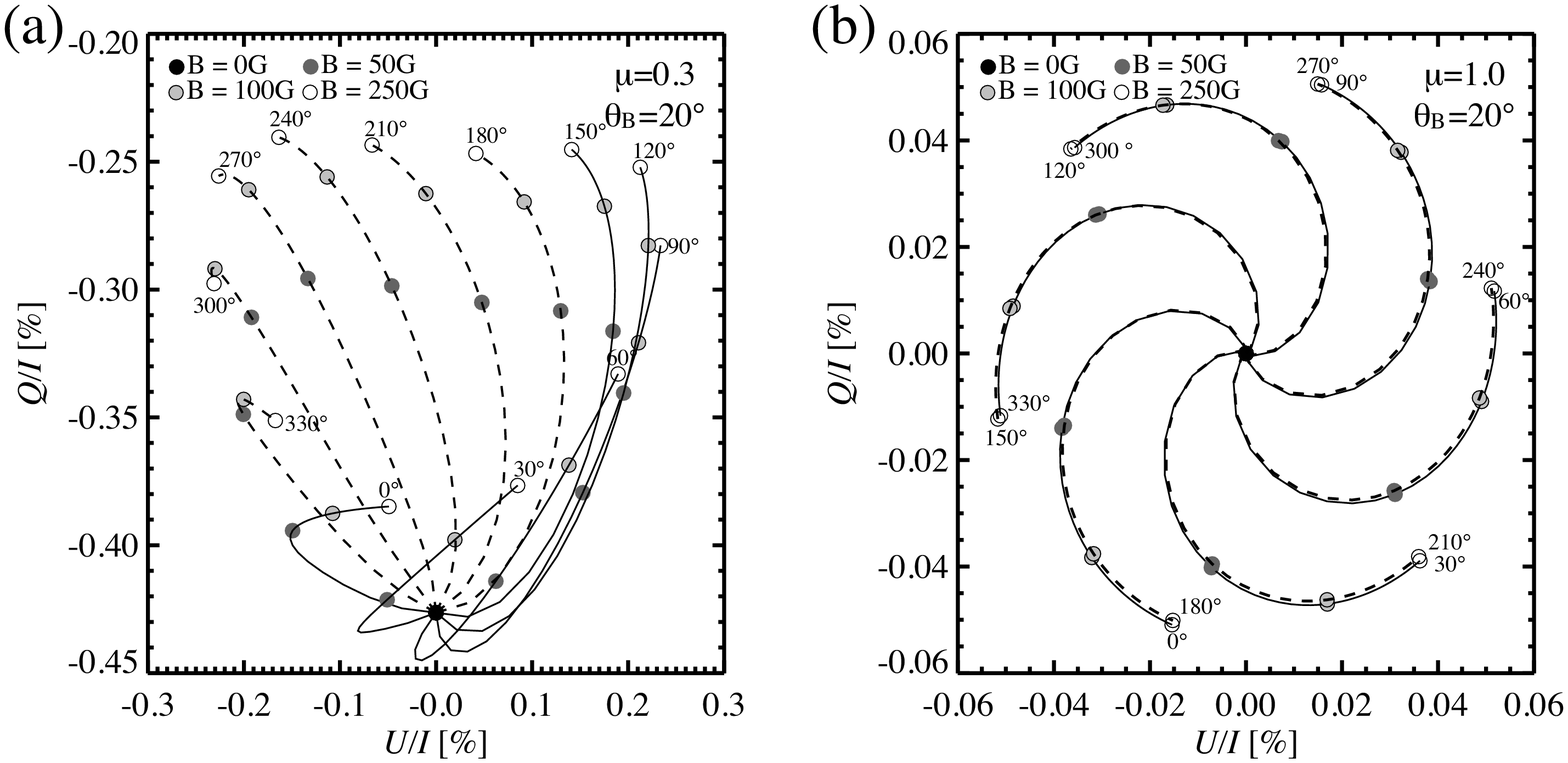}
\caption{Hanle diagrams for the case of a nearly vertical field ($\theta_B=20^{\circ}$).
The curves were generated in the same way as those in Figure~\ref{fig:diagram1}.
}
\label{fig:diagram2}
\end{figure}

The behavior of the $Q/I$ and $U/I$ signals
with respect to the magnetic field parameters (strength, inclination, and azimuth of the magnetic field vector)
can be suitably illustrated by a Hanle diagram \citep[see also][]{Trujillo2012}.
In Figure~\ref{fig:diagram1},
we show Hanle diagrams with a horizontal magnetic field (i.e.,
the inclination is fixed at $\theta_B=90^{\circ}$)
for different values of the azimuth angle when
the field strength varies between 0~G (black circles) and 250~G (white circles).
The Figure~\ref{fig:diagram1}~(a) corresponds to a close-to-limb geometry ($\mu=0.3$), while Figure~\ref{fig:diagram1}~(b)
refers to the forward scattering case of a disk center observation ($\mu=1.0$).
The $Q/I$ and $U/I$ signals in this figure
refer to the amplitudes of original synthetic profiles
$Q(\lambda_0)/I(\lambda_0)$
and $U(\lambda_0)/I(\lambda_0)$ in the database
at the line center, where $\lambda_0=121.567$~nm.
Figure~\ref{fig:diagram2} is similar to Figure~\ref{fig:diagram1},
but the magnetic field is nearly vertical (i.e., the inclination
is fixed at $\theta_B=20^{\circ}$).

For the close-to-limb geometry ($\mu=0.3$),
the unmagnetized case ($B=0$~G) shown in black circles in Figs.~\ref{fig:diagram1}~(a) and
\ref{fig:diagram2}~(a) yields negative Stokes $Q/I$ values, which indicates that the direction
of linear polarization caused by the anisotropic radiation field
(i.e., the non-magnetic scattering polarization)
is perpendicular to the solar limb.
As shown by \citet{Trujillo2011}, the anisotropy of
the radiation field, $J^2_0$,
illuminating the hydrogen atoms in the Lyman-$\alpha$ line
is negative (dominated by horizontal illumination)
through the line formation region
of the FAL-C model atmosphere. The
resulting scattering polarization is perpendicular to the limb.
At the disk center ($\mu=1.0$), where
the line of sight (LOS) is parallel to the solar normal (i.e., the symmetry axis of the radiation field),
we have $Q/I=0$ and $U/I=0$
for the non-magnetized case (black circles in Figs.~\ref{fig:diagram1}~(b) and \ref{fig:diagram2}~(b)).

The crossing points seen in the $\mu=0.3$ Hanle diagrams
indicate ambiguity, which occurs when different magnetic field vectors
give the same Stokes $Q/I$ and $U/I$ signals.
For example, in Figure~\ref{fig:diagram1}~(a),
the crossing point at $Q/I\sim-0.3\%$ and $U/I\sim-0.15\%$
corresponds to cases with $\chi_B=330^{\circ}$ and $B=10$~G
and with $\chi_B=60^{\circ}$ and $B=30$~G.
This ambiguity cannot be solved without using additional information
to constrain one of magnetic parameters.
At the solar disk center ($\mu=1.0$), we have
$180^{\circ}$ ambiguity, where two magnetic field vectors
whose azimuths differ by $180^{\circ}$ represent the same $Q/I$ and $U/I$ signals (see overlapping solid and dashed lines
in Figs.~\ref{fig:diagram1}~(b) and \ref{fig:diagram2}~(b)).

In such Hanle diagrams, the change in linear polarization from $0$~G to $50$~G
is larger than that from $50$~G to $250$~G, with the exception of
nearly vertical fields in the $\mu=1.0$ forward scattering geometry case
(Figure~\ref{fig:diagram2}~(b)).
This trend is prominent for the horizontal field case
shown in Figure~\ref{fig:diagram1},
where $Q/I$ and $U/I$ significantly change from 0 to 50~G but
show little change from 50~G to 250~G.
This indicates that the Hanle effect in the hydrogen Lyman-$\alpha$
line is sensitive to field strengths $B{<}50$~G.
A field strength of $\sim50$~G corresponds to the
critical field strength for the onset of the Hanle effect in
the Lyman-$\alpha$ line.
Above this field strength, the Lyman-$\alpha$ line approaches the Hanle saturation limit,
where the linear polarization is insensitive to the magnetic field strength.

The linear polarization signals produced by the Hanle effect are weaker than 0.1\%
for a nearly vertical field
at the solar disk center ($\mu=1.0$)
(Figure~\ref{fig:diagram2}~(b))
because weak vertical fields
do not give rise to a strong symmetry breaking.
At the solar disk center, largely inclined and strong fields
are more suitable observing targets.

\section{Inversion}
\subsection{General approach for solving the inversion problem}
\label{chidef}
\begin{figure}
\epsscale{1.0}
\plotone{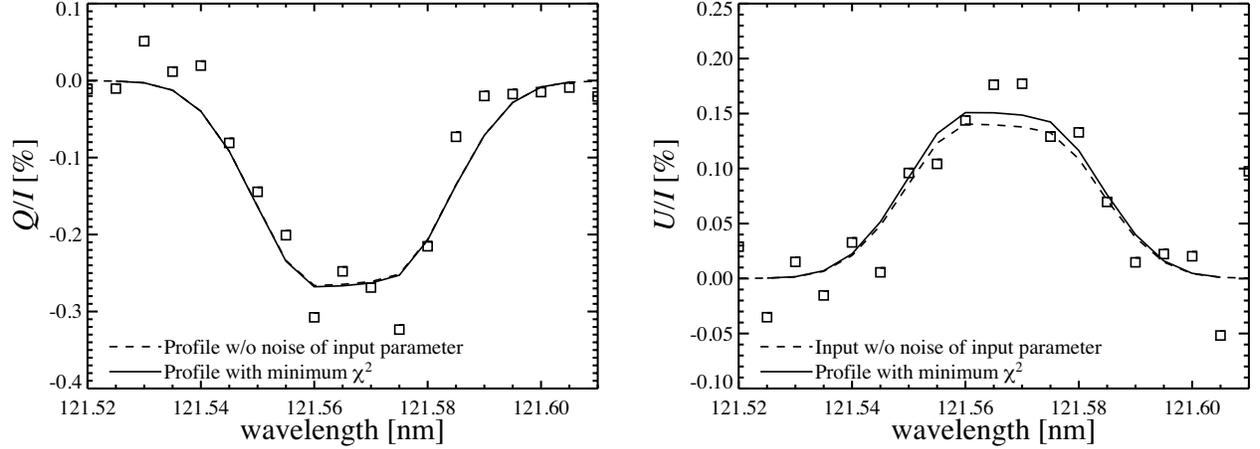}
\caption{The dashed lines show the convolved synthetic profiles for input with the magnetic parameters
of $B=50$~G, $\theta_B=90^{\circ}$, and
$\chi_B=120^{\circ}$.
The simulation of the observed profiles (noise with $\sigma=0.033\%$ is added into the input profiles) is shown with squares.
The solid lines show the convolved synthetic profiles
for one of acceptable solutions with numerically minimal $\chi^{2}$ values (shown with the gray triangle in Figure~\ref{fig:3dchi}).
}
\label{fig:invprof}
\end{figure}

To clarify whether we can retrieve information on the magnetic field from the observed Stokes profiles,
we perform a process that mimics the Stokes inversion.
Here, we assume that the formation of the Lyman-$\alpha$ line is modeled with a semi-empirical
FAL-C model atmosphere.
For this purpose, we introduce the following function:
\begin{equation}
\chi^{2}(B,\chi_B,\theta_B)\equiv\sum_{k=1}^{2}\sum_{l=1}^{n}
\frac{[S_{k}^{obs}(\lambda_{l})-S_{k}^{mod}(\lambda_{l},B,\chi_B,\theta_B)]^{2}}{\sigma^2},
\end{equation}
where $S_{1}$ and $S_{2}$ indicate $Q/I$ and $U/I$, respectively.
The simulated CLASP observation is $S_{k}^{obs}$ ($k=1,2$),
i.e., the synthetic Stokes profile taken
from the database, convolved with a 0.013~nm FWHM Gaussian,
sampled with a wavelength step of 0.005~nm, with added noise
(Section~\ref{database}).
Here, $S_{k}^{mod}$ represents the noiseless  model profile corresponding to
the synthetic profile convolved with a 0.013~nm FWHM Gaussian
and sampled with a wavelength step of 0.005~nm.
In the database, the parameter increments in the synthetic profiles are
5~G increments of field strength and 5$^{\circ}$ increments
of azimuth and inclination.  In order to have better accuracy in
the inversion, $S_{k}^{mod}$ are calculated with
$1^{\circ}$ steps for azimuth and inclination and
with $1$~G steps for field strength
by linearly interpolating
model profiles with adjacent magnetic parameters.
We take into account the wavelength range
of $121.567~\mathrm{nm}\pm0.04~\mathrm{nm}$
where the linear polarization signals are defined,
resulting in $n=17$ wavelength points.
Finally, $\sigma$ is the standard deviation of the random noise we assumed
in the CLASP observation simultations.
We employ $\sigma=0.033\%$ as the baseline, and in Section~\ref{noise},
we will discuss the influence of noise on the inversion results.

We calculate the $\chi^{2}$ function
using the given observed profile ($S_{k}^{obs}$) and all of the model profiles
($S_{k}^{mod}$).  We find the magnetic parameters with a statistically acceptable $\chi^{2}$,
defined by
$\Delta\chi^2\equiv\chi^2-\chi^2_{\mathrm{min}}\le3.53$,
where $\chi^2_{\mathrm{min}}$ is the minimum $\chi^2$ \citep{Numerical}.
Because we have three free parameters ($B$, $\chi_B$, and $\theta_B$),
$\Delta\chi^2\equiv\chi^2-\chi^2_{\mathrm{min}}\le3.53$
is given by the chi-square distribution
function for three degrees of freedom with a confidence level of 68.3\% ($1\sigma$).
The parameter region defined by this criteria indicates that there
is a 68.3\% ($1\sigma$) chance for the true
field strength, azimuth, and inclination parameters to fall within this
region.
We call this procedure ``inversion'' throughout this paper.
Furthermore, we use the term ``input'' to refer to
the magnetic parameters of the simulated observation $S_{k}^{obs}$.

\begin{figure}
\epsscale{1.0}
\plotone{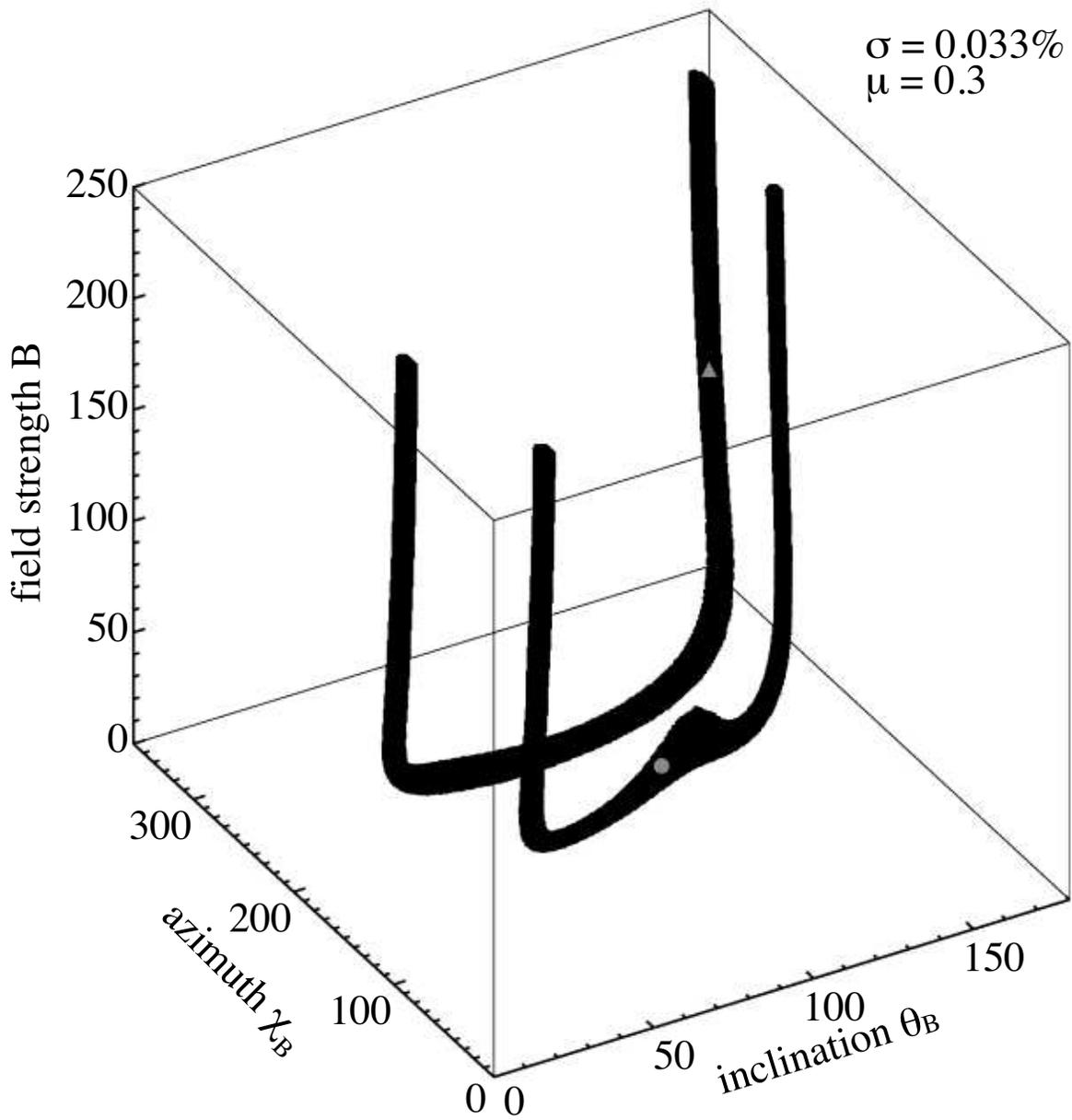}
\caption{The $\chi^2$ map in the field strength, azimuth, and inclination parameter space.
The magnetic parameters with $\Delta\chi^2\le3.53$ (see Section~\ref{chidef})
are shown in black.
The input magnetic parameters
of $B=50$~G, $\theta_B=90^{\circ}$, and $\chi_B=120^{\circ}$
are shown with a gray circle.
The gray triangle shows one of the
acceptable solutions (magnetic parameter with
numerically minimal $\chi^{2}$).}
\label{fig:3dchi}
\end{figure}

Figure~\ref{fig:3dchi} shows the dependence of $\chi^2$ on inclination, azimuth, and field strength
for the input parameters of $B=50$~G, $\theta_B=90^{\circ}$, and
$\chi_B=120^{\circ}$, which are shown by gray circles.
The black region in Figure~\ref{fig:3dchi} is defined by
$\Delta\chi^2\equiv\chi^2-\chi^2_{\mathrm{min}}\le3.53$.
The model profiles located in this region fit the simultated observations reasonably well,
and all magnetic parameters are statistically accepted as results
of inversion with a confidence level of 68\% ($1\sigma$).
The model profiles corresponding to the input parameters
are shown with dashed lines in Figure~\ref{fig:invprof}.
The magnetic parameters
(chosen as an example in the acceptable $\chi^2$ region)
marked by gray triangle in Figure~\ref{fig:3dchi}
correspond to the model profiles
shown with solid lines in Figure~\ref{fig:invprof}.
Within the noise level, these profiles are identical.

\begin{figure}
\epsscale{1.0}
\plotone{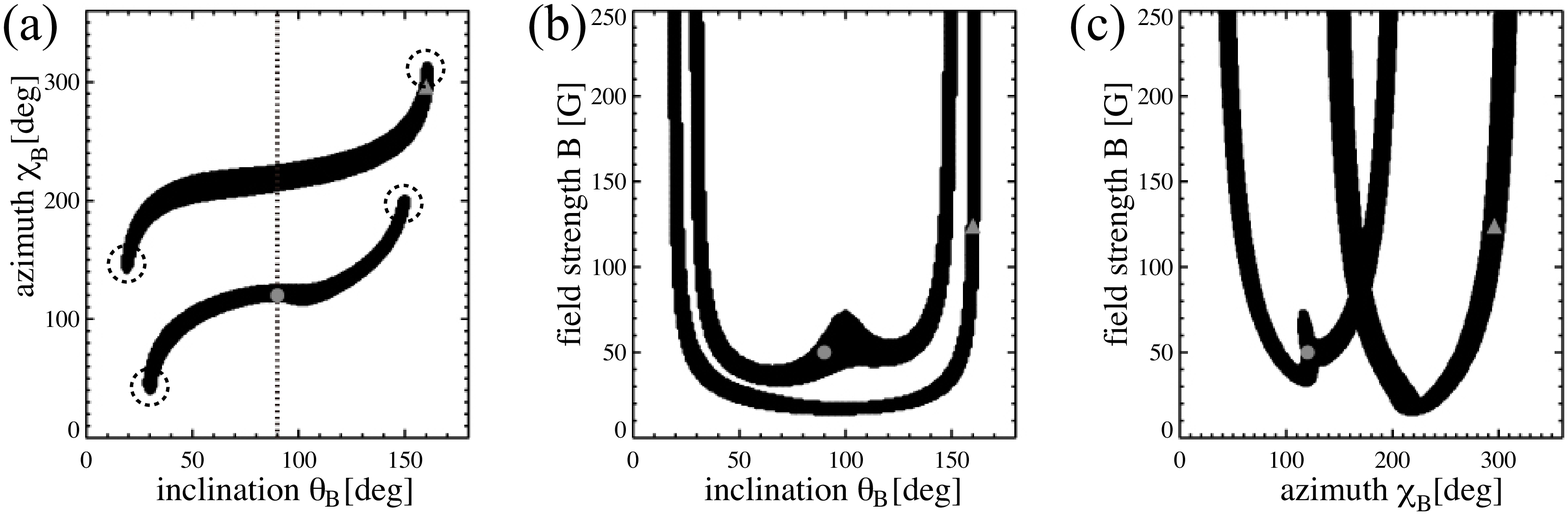}
\caption{The $\chi^2$ maps projected onto the parameter space of (a) inclination and azimuth, (b) inclination and field strength, and (c) azimuth and field strength
for the input parameters of $B=50$~G, $\theta_B=90^{\circ}$, and $\chi_B=120^{\circ}$.
The regions with $\Delta\chi^{2}\le3.53$ are shown in black.
The input magnetic parameters are shown with gray circles.
The gray triangle represents one of acceptable solutions shown in
Figure~\ref{fig:3dchi}. The four dashed circles in panel (a) indicate the four ambiguous solutions in saturation regime.}
\label{fig:2dchi}
\end{figure}

\subsection{Properties of the $\chi^2$ map}
\label{chi2property}
\subsubsection{Saturation regime}
\label{saturation}

Figure~\ref{fig:2dchi} represents the $\chi^2$ maps projected onto
two parameter spaces.
There are four 1$\sigma$ confidence level regions ($\Delta\chi^2\le3.53$)
extended along the field strength above 50~G
at $\theta_B=20^{\circ}$, $30^{\circ}$, $150^{\circ}$, $155^{\circ}$
in Figure~\ref{fig:2dchi}~(b) and
at $\chi_B=40^{\circ}$, $145^{\circ}$, $195^{\circ}$, $310^{\circ}$
in Figure~\ref{fig:2dchi}~(c).
The four dashed circles in Figure~\ref{fig:2dchi}~(a)
show the regions
on the $\theta_B$-$\chi_B$ plane
where $(\theta_B, \chi_B)=\{(20^{\circ}, 145^{\circ}),
(30^{\circ}, 40^{\circ}), (150^{\circ}, 195^{\circ}),
(155^{\circ}, 310^{\circ})\}$,
indicating four possible solutions
of inclination and azimuth above 50~G.
The elongated line, which occurs in only field strength direction,
indicates a large \textit{uncertainty} in the field strength
where we cannot constrain the field strength.
In other words, we do not have sensitivity to measure
the magnetic field strength beyond $\sim50$~G.
Multiple simply-connected spaces
($1\sigma$ confidence level regions) indicate
the \textit{ambiguity} of solutions in which completely
different magnetic parameters
provide the same $Q/I$ and $U/I$ profiles.

The elongated regions correspond to the saturation regime,
in which the linear polarization weakly depends on the field strength.
This is consistent with the Hanle diagrams
(Section~\ref{hanle_diagram}), where
the change of linear polarization above 50~G is small compared
with that below 50~G.
In this saturation regime, the linear polarization signal is only dependent
on the inclination and azimuth.
Thus, we can determine the azimuth and inclination.
However, ambiguity allows
four combinations of inclination and azimuth to exist
as shown in Figure~\ref{fig:2dchi}.
The ambiguity corresponds to the Van Vleck ambiguity
in the saturation regime, which is inherent to the Hanle effect.
These four ambiguous solutions provide magnetic parameters with less
inclined ($\theta_B\sim20-30^{\circ}$ or $\theta_B\sim150-160^{\circ}$)
and relatively strong ($B>50$~G) magnetic field vectors.

\subsubsection{Non-saturation regime}
\label{non-saturation}

Below 50~G, the linear polarization of $Q/I$ and $U/I$
depends on the field strength as well as on the inclination and
azimuth (Section~\ref{hanle_diagram}); this is the non-saturation regime.
As shown in Figure~\ref{fig:2dchi}~(a),
in the non-saturation regime,
there are two isolated $1\sigma$ confidence level regions
with saturation regimes at both ends (dashed circles, four in total).
These two isolated regions are elongated over the wide
inclination and azimuth range on the $\theta_B$-$\chi_B$ plane.
Different magnetic parameters give rise to the
same linear polarization signals in these two regions.
This can be confirmed by the Hanle diagram in 
Figure~\ref{fig:diagram1}~(a).
If we assume that the magnetic field is horizontal ($\theta_B=90^{\circ}$),
as shown in the dashed line in Figure~\ref{fig:2dchi}~(a),
two solutions are possible:
one for $\chi_B=120^{\circ}$ and $B=50$~G (shown in gray circle)
and the other for $\chi_B=225^{\circ}$ and $B=15$~G.
Indeed, the Hanle diagram with $\chi_B=120^{\circ}$
and $B=50$~G intersects with that with $\chi_B=225^{\circ}$ and $B=15$~G,
suggesting the presence of ambiguity.

Below $\sim50$~G, Figure~\ref{fig:2dchi}~(b) shows
that one $1\sigma$ confidence level region, which extends
from $\theta_B\sim20^{\circ}$ to
$\theta_B\sim150^{\circ}$,
has an apex at $B\sim15$~G and $\theta_B\sim90^{\circ}$.
Figure~\ref{fig:2dchi}~(b) further shows that another $1\sigma$ confidence level region,
which extends in the
$\theta_B=30^{\circ}$ to $140^{\circ}$ range,
possesses two vertices
at $\theta_B\sim60^{\circ}$ and $\theta_B\sim120^{\circ}$.
The elongated shape over the inclination and field strength
shows a strong \textit{correlation}
between these two magnetic parameters;
both a weak and inclined field
and a stronger and less inclined field provide equally good fitting
to the observed spectra.
On the $\chi_B$-$B$ plane,
the two $1\sigma$ confidence level regions
have a V-shape with apexes at
$\chi_B\sim120^{\circ}$
and $B\sim40$~G and at $\chi_B\sim220^{\circ}$
and $B\sim15$~G.
This also suggests that there is a correlation between
the azimuth and field strength and that the
wide field strength range is consistent with the data.
We notice that the azimuth converges to $120^{\circ}$ or $220^{\circ}$ when
the field strength becomes weaker.
In general, there is strong correlation among
these three magnetic parameters, and it is difficult
to uniquely determine a set of these magnetic parameters.

\subsubsection{Connection between saturation and non-saturation regimes}

We find four-fold ambiguity in the saturation regime ($B>50$~G) and
two-fold ambiguity in the non-saturation regime ($B<50$~G).
As we clearly show in Figure~\ref{fig:3dchi},
a pair of saturated regimes converges into one of the non-saturated regimes
(there are two sets of connections).
This connectivity indicates that the Stokes $Q/I$ and $U/I$ profiles
for a strong and less inclined magnetic field
is similar to those of a weaker and more inclined field.
This transition suggests that the
degree of the Hanle effect remains the same
among strong, less inclined fields and weaker, more inclined fields.
In summary, multiple solutions are possible over a broad field strength range.
In the saturation regime ($B>50$~G),
we can only determine the azimuth and inclination, although we have
multiple solutions in these parameters.
As we enter the non-saturation regime, we have strong
correlations among three magnetic parameters in addition to the above
ambiguity.
These situations make it difficult
to uniquely determine the magnetic field vector.

\subsection{Additional information for constraining magnetic parameters}
\label{mag_constrain}
If there are multiple solutions, one way to uniquely determine the magnetic field vector
is to constrain one of parameters using additional observations.
With the exception of the saturation regime,
the 1$\sigma$ confidence level regions are extended over
the three-dimensional parameter space as
shown in the 2D $\chi^{2}$ maps in Figure~\ref{fig:2dchi}.
The shape helps us to constrain the magnetic parameters
with an additional piece of information.
For example, once the azimuth is constrained by other observations,
the inclination and field strength will be uniquely determined
as inferred from Figure~\ref{fig:2dchi}~(a) and (c).
An accuracy of $\pm5^{\circ}$ in the azimuthal direction will
be good enough for most cases.
However, when $\chi_B=215^{\circ}$ and $\chi_B=120^{\circ}$,
the correlation curve on the plane of inclination and azimuth
allows a relatively large uncertainty in the inclination (see Fig.\ref{fig:2dchi}~(a)).

The Lyman-$\alpha$ images of the upper chromosphere obtained with the Very high Angular resolution ULtraviolet Telescope (VAULT) sounding rocket \citep{Vourlidas2010}
show the presence of long thin threads of $\sim10\arcsec$ in the quiet Sun.
The Mg II k and Ca II images
obtained with the Sunrise FilterImager \citep[SuFI;][]{Grandorfer2011}
revealed fibril structures spreading from the plage regions
in the chromosphere \citep{Riethmuller2013}.
As shown by \citet{Leenarts2012,Leenarts2013},
the magnetic field connecting magnetic concentrations with opposite polarities represents the intensity filamentary structure in their 3D atmospheric model.
This indicates that the intensity filamentary structures can be used as proxies of the azimuth of the magnetic fields.
Therefore, high-spatial resolution observations around the
line forming layer of the Lyman-$\alpha$ line
would help us to know the azimuthal direction.
The CLASP Lyman-$\alpha$ slit-jaw images, 
the Interface Region Imaging Spectrograph (IRIS),
the Atmospheric Imaging Assembly (AIA) onboard the Solar Dynamics Observatory (SDO), and ground-based observations can be used for this purpose.

\subsection{Inversion with different noise levels}

\label{noise}
\begin{figure}
\epsscale{1.0}
\plotone{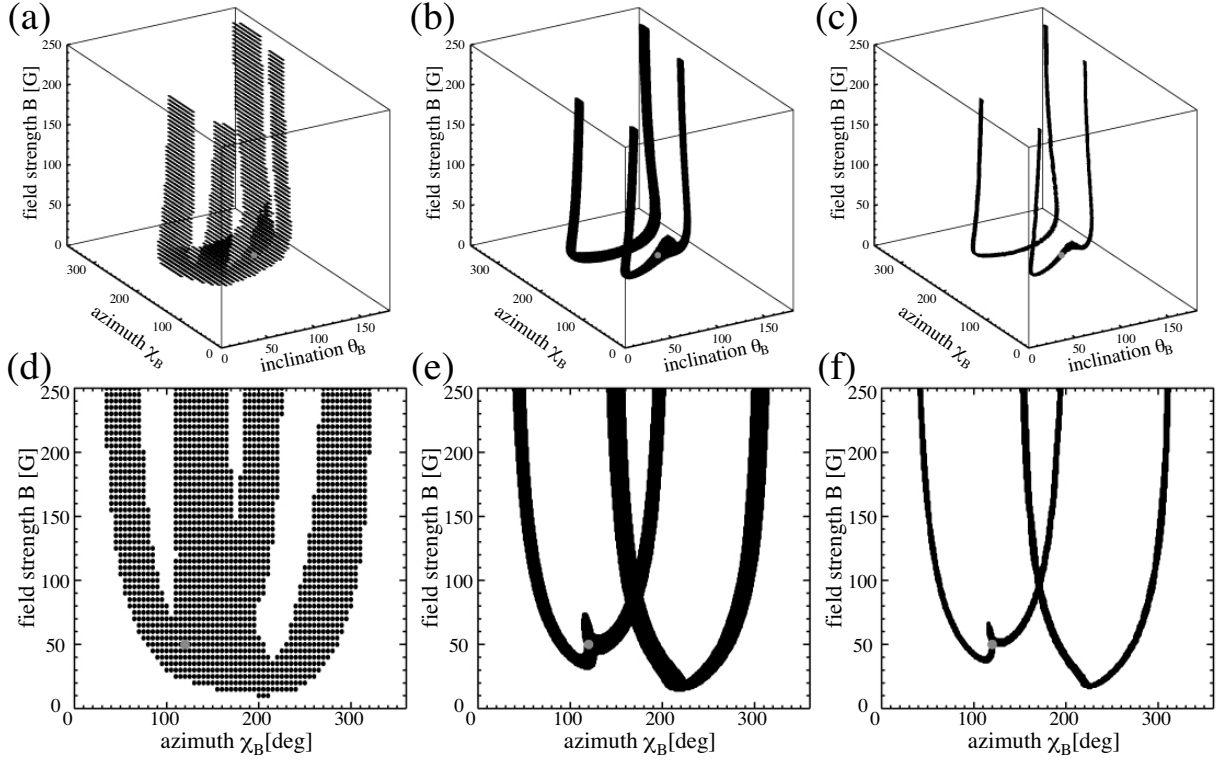}
\caption{
The $\chi^2$ maps in the field strength, azimuth, and inclination parameter space
with the input parameters of $B=50$~G, $\theta_B=90^{\circ}$, and $\chi_B=120^{\circ}$ at $\mu=0.3$
are shown for three different noise levels: (a) $1\sigma=0.1\%$,
(b) $1\sigma=0.03\%$, and (c) $1\sigma=0.01\%$.
(d), (e), and (f) are the $\chi^{2}$ maps from (a), (b), and (c), respectively, projected onto the azimuth and field strength parameter space.
In (a) and (d), the regions with $\Delta\chi^{2}\le3.53$ are
plotted with $5^{\circ}$ increments in $\chi_B$ and $\theta_B$ and with 5~G increments in $B$.
The input magnetic parameters are shown with gray circles in all panels.}
\label{fig:noise}
\end{figure}

Here, we investigate the influence of the noise level on the inversion results.
Figure~\ref{fig:noise} represents the dependence of the $\chi^{2}$ values
on inclination, azimuth, and field strength
for noise levels of $1\sigma=0.1\%$,
$1\sigma=0.03\%$, and $1\sigma=0.01\%$.
The input parameters for all cases are
$B=50$~G, $\theta_B=90^{\circ}$, and $\chi_B=120^{\circ}$.
Note that Figure~\ref{fig:noise}~(b) is the same as Figure~\ref{fig:3dchi}.
The $\chi^{2}$ distributions are similar
for all noise levels. Above 50~G, four isolated $1\sigma$ confidence regions
with $\Delta \chi^{2}\le3.53$
are extended only in the direction of
the field strength, representing the saturation regime. Around 50~G,
these regions are connected to the two isolated regions
with largely inclined, weak fields.
Even though we increase the signal-to-noise ratio,
the ambiguity remains intact,
and we find similar correlation between the magnetic parameters.
Thus, we require additional observables to constrain
the solution even in low noise situations.
The difference caused by different noise levels is equivalent to the thickness of
the $1\sigma$ confidence level region.
These regions are thinner for lower noise levels, indicating
lower uncertainty in the determination of the magnetic parameters.
For example, in the saturation regime,
the thicknesses correspond to $40^{\circ}$ for $1\sigma=0.1\%$,
$\sim$10$^{\circ}$ for $1\sigma=0.033\%$, and $\sim$5$^{\circ}$
for $1\sigma=0.01\%$
(see Figure~\ref{fig:noise}~(d), (e) and (f)).

\subsection{Inversion for different observing regions and input parameters}

To see whether the properties of the $\chi^{2}$ maps (i.e., result of inversion)
identified in Section \ref{chi2property} depend
on the choice of the input parameters,
we perform inversions for different sets of input parameters and for
different observing locations on the solar disk.
We study input parameters with weak field ($B=10$~G),
marginal field ($B=50$~G),
and strong field ($B=250$~G) with both horizontal ($\theta_B=90^{\circ}$) and
almost vertical ($\theta_B=20^{\circ}$) configurations.
Note that all azimuths are fixed at $\chi_B=120^{\circ}$.
With this set of input parameters, we consider both the close-to-limb case ($\mu=0.3$)
and the disk center case ($\mu=1.0$) and execute 12 total inversions.

\begin{figure}
\epsscale{0.8}
\plotone{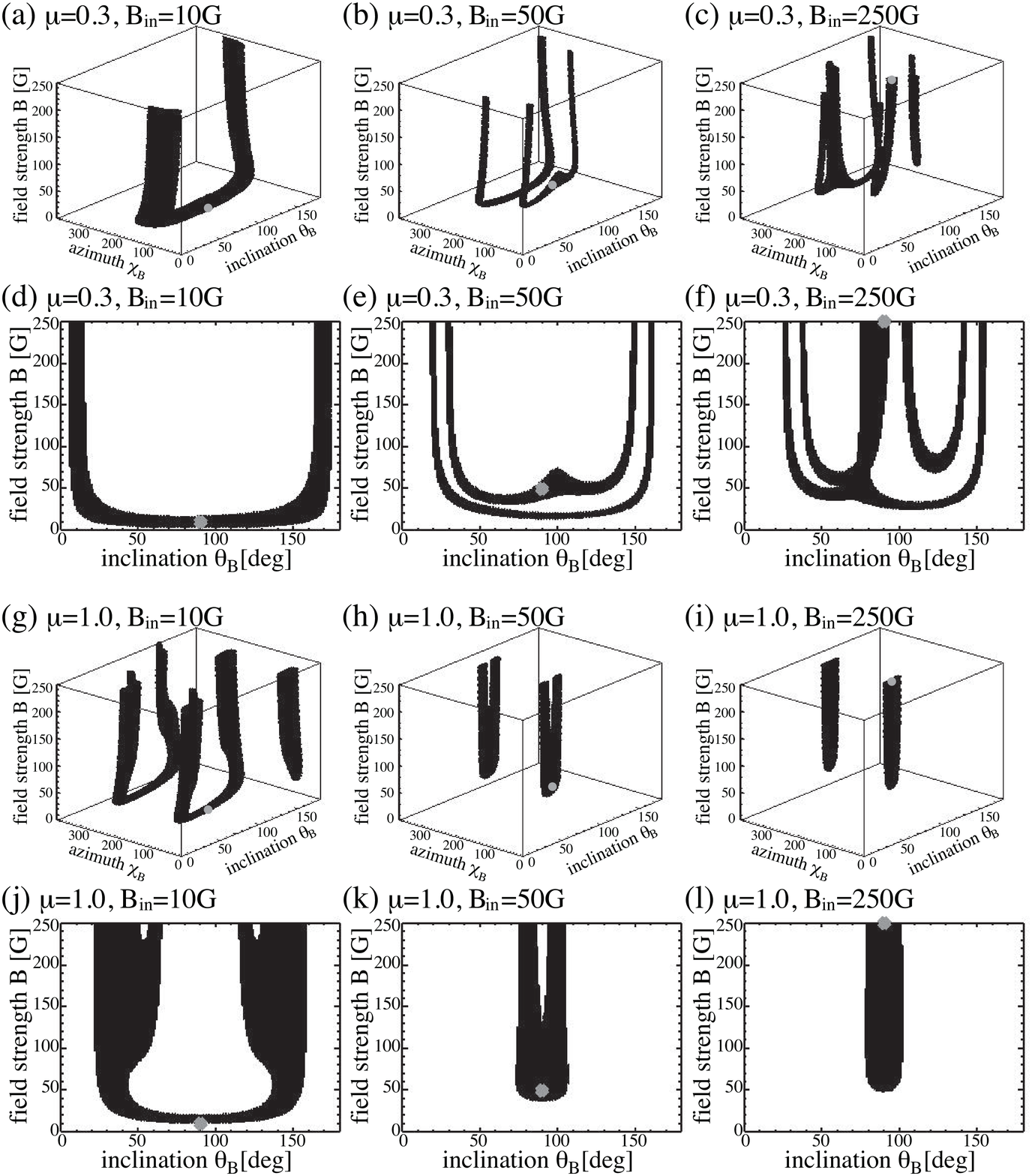}
\caption{Plots (a), (b), and (c), and plots (g), (h), and (i) represent the
$\chi^2$ maps in the field strength, azimuth, and inclination parameter space
for a horizontal magnetic field ($\theta_B=90^{\circ}$) input
with the close-to-limb ($\mu=0.3$) and the disk center ($\mu=1.0$) geometries,
respectively.
Plots (d), (e), and (f) and
plots (j), (k), and (l) represent the $\chi^{2}$ maps
projected onto the inclination and field strength parameter space.
The left, middle, and right columns show $\chi^2$ maps
with a noise level of $\sigma=0.033\%$ for
the inputs of
weak field ($B=10$~G),
marginal field ($B=50$~G), and strong field ($B=250$~G), respectively,
The black regions represent the magnetic parameters with $\Delta\chi^2\le 3.53$.
The input magnetic parameters are shown with gray circles.}
\label{fig:3dchi_horizontal}
\end{figure}

Figure~\ref{fig:3dchi_horizontal} shows $\chi^2$ maps for the case of
a horizontal magnetic field. With the exception of the disk center case ($\mu=1.0$)
with $B=50$~G (Figure~\ref{fig:3dchi_horizontal}~(h) and (k))
and $B=250$~G (Figure~\ref{fig:3dchi_horizontal}~(i) and (l)),
the $\chi^2$ distributions
are similar to those in Section~\ref{chi2property}, and
any field strength, from weak to strong, is consistent with the data.
For the disk center, with horizontal magnetic fields of 50 and 250~G,
the $1\sigma$ confidence level region appears only in the saturation
regime. Thus, in this case, there is no possibility to
have a wrong solution with weak magnetic fields.
This is distinct advantage; however, is impossible to determine the field strength in this case.
The number of ambiguous regions is different depending on the input parameters.

\begin{figure}
\epsscale{0.8}
\plotone{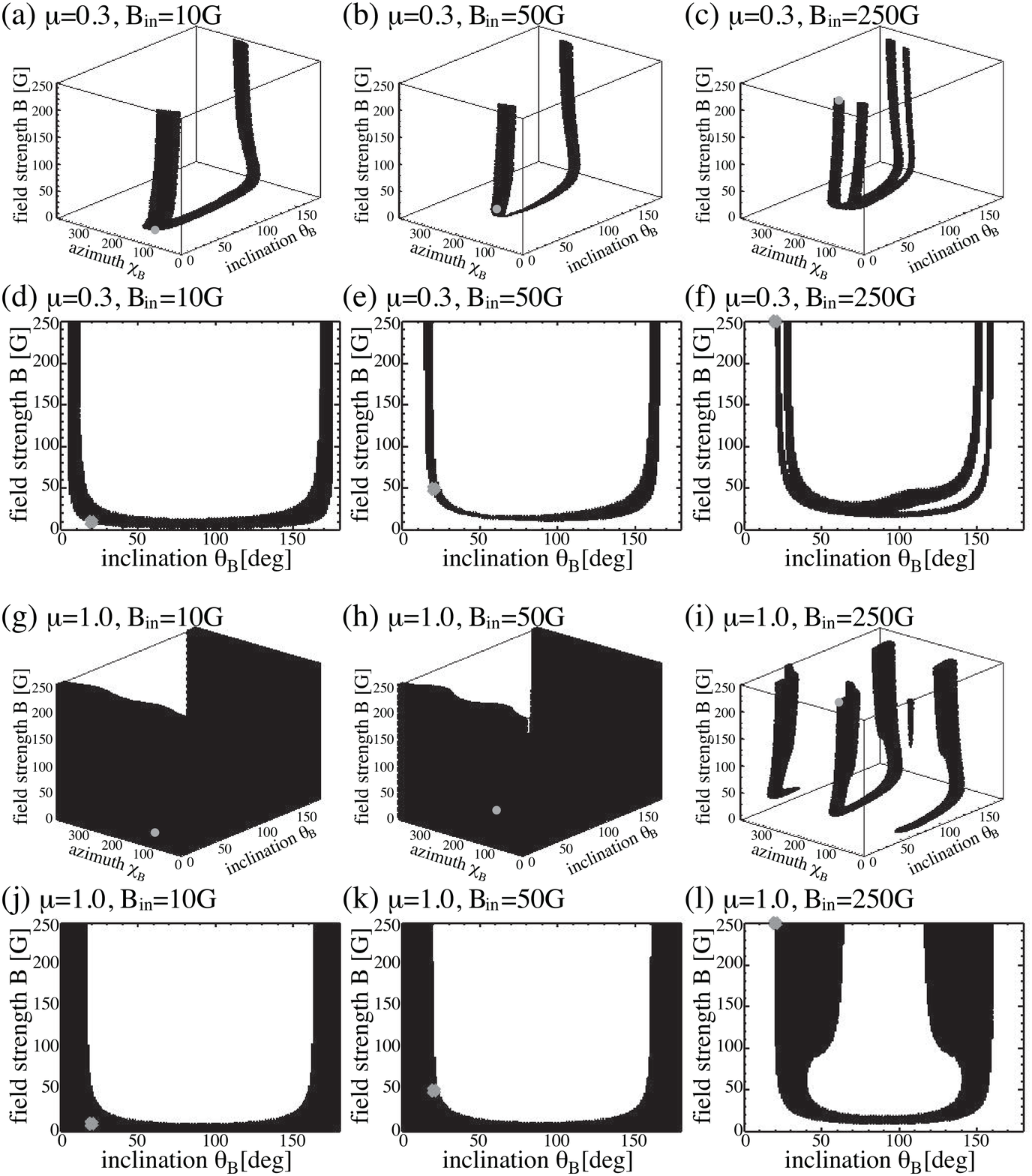}
\caption{The plots shown here are the same as those in Figure~\ref{fig:3dchi_horizontal}
with an almost vertical magnetic field ($\theta_B=20^{\circ}$) input value.}
\label{fig:3dchi_vertical}
\end{figure}

Figure~\ref{fig:3dchi_vertical} shows $\chi^2$ maps for the case
of vertical magnetic fields ($\theta_B=20^{\circ}$).
Again, we find the same properties of $\chi^2$ distributions, indicating that any field strength can be possible as a solution.
For the disk center case with $B=10$~G and $B=50$~G
(Figure~\ref{fig:3dchi_vertical}~(g) and (h)),
the $1\sigma$ confidence regions with $\Delta\chi^{2}\le3.53$ spread out
on the plane with vertical fields ($\theta_B=0^{\circ}$ or $\theta_B=180^{\circ}$)
and on the plane with zero magnetic fields
(see Figure~\ref{fig:3dchi_vertical} (j) and (k)),
indicating that the inversion does not work.
This result is reasonable because the
the magnitude of linear polarization is quite small and below the noise level
when the magnetic fields
are almost vertical, as shown in Figure~\ref{fig:diagram2}~(b).
If we employ a smaller noise level,
properties of the $\chi^{2}$ distribution similar to those in
Section~\ref{chi2property} will appear.
For CLASP observations, an observing target
with a largely inclined and/or strong
magnetic field strength is appropriate
for the disk center observation.

\section{Discussions}
\subsection{Required additional information}

We have performed Stokes inversion simulations to clarify
the information which can be inferred
via the Hanle effect in the hydrogen Lyman-$\alpha$ line, 
assuming that the chromosphere and transition region of the quiet Sun 
can be represented by the FAL-C semi-empirical model.
We conclude that UV spectro-polarimetry with the CLASP experiment
is  a suitable diagnostic tool of the magnetic field
in the upper atmosphere, if combined with complementary information from other relevant  observations.
Though we have the ambiguity and uncertainty
that is inherent to the Hanle effect when only the scattering polarization in 
one spectral line is available, 
this should not be taken as a drawback.
As we have shown, we need additional observations to uniquely determine the
field strength, azimuth, and inclination. Clearly, we cannot measure the 
very small contribution of the Zeeman effect to the Stokes $V$ of the Lyman-$\alpha$ line, but there
are several options for resolving this issue.
Ideally, we would like to perform
simultaneous spectro-polarimeteric observations also 
in other spectral lines of the upper chromosphere, 
which have different critical field strengths for the onset of
the Hanle effect \citep{Trujillo2012,Belluzzi2012}. 
However, in this paper, we propose a simpler,
but useful, third method for determining
the azimuthal magnetic field direction   
using the fibrils seen in the high-resolution intensity images 
from IRIS, AIA, and ground-based observations. 

\subsection{Observing target}

The Lyman-$\alpha$ line starts to approach 
the Hanle saturation regime above $\sim$50~G,
where the linear polarization changes only with the inclination and azimuth of the magnetic field, 
not with its strength.
Furthermore, nearly vertical fields do not produce any significant  
Hanle effect (i.e., the magnetic modification of the linear polarization), and at the solar 
disk center the linear polarization created by the Hanle effect of 
slightly inclined fields is too small to be detected.
Thus, inclined, relatively weak ($B<50$~G) magnetic fields should be observed.
Based on the properties of the Hanle effect studied in this paper,
we can now discuss the possible observing region and observing target for the CLASP experiment.

Our primary goal with the CLASP experiment
is to detect for the first time the linear polarization
caused by the atomic level polarization produced  
by the absorption and scattering of anisotropic radiation in the
upper solar atmosphere.
To this end, it is desirable to choose a quiet region close to the limb 
(e.g., around $\mu{\approx}0.3$) because such locations are the most suitable ones for
detecting the line-core polarization in the hydrogen Lyman-$\alpha$ line \citep{Trujillo2011,Belluzzi2012b,Stepan2014}. 
Our second goal is to detect the Hanle effect, in order to constrain
the magnetic field vector of the chromosphere-corona transition region.

One of the popular spectral lines for magnetic field measurements 
in the upper atmosphere is the He~{\sc i} 1083~nm triplet \citep[e.g.,][]{Asensio2008}.
By exploiting the spectro-polarimetric data obtained with this multiplet,
the magnetic properties of prominences, filaments, spicules 
and active regions have been investigated by several authors \citep[e.g.,][]{Trujillo2002,Lagg2004,Marenda2006,Centeno2010,Xu2010}.
However, it is not easy to measure the intensity and polarization of the He~{\sc i} 1083 nm triplet 
in quiet regions of the solar disk \citep[e.g.,][]{Asensio2008},
and there are few studies on the quiet-Sun magnetic fields of the upper solar 
atmosphere. Thus, our primary targets are the network and internetwork regions
of the quiet Sun. The network fields are expected to
form magnetic canopy structures in the upper chromosphere and transition region,
and they are expected to be largely inclined and relatively weak.
\citet{Wiegelmann2010} investigated the fine structure of the magnetic fields
in the quiet Sun using photospheric magnetic field
measurements from the SUNRISE imaging magnetograph experiment (IMaX).
\citet{Wiegelmann2010} found that most magnetic loops rooted in the quiet Sun photosphere
would reach into the chromosphere or higher.
In addition to the canopy field, such magnetic loops in regions of the quiet Sun  
would also be interesting observing targets.

\subsection{Atmospheric model}

Finally, we discuss another issue 
that we should address further in future investigations: 
the influence of the atmospheric model on the inference of the 
magnetic field via the interpretation of the scattering polarization 
and the Hanle effect in Lyman-$\alpha$.
\citet{Belluzzi2012b} calculated the scattering polarization
profiles of the hydrogen Lyman-$\alpha$ line taking into account partial frequency redistribution
(PRD) and $J-$state interference effects, and using the plane-parallel atmospheric models C, F, and P of \citet{Fontenla1993}, which can be considered as illustrative of quiet, network, and plage
regions. 
They showed that the shape and amplitude of the Lyman-$\alpha$  
linear polarization profiles are sensitive
to the thermal structure of the model atmosphere
in the line wings, and to a lesser extent also in the line core (where the Hanle
effect operates). 
Thus, in order to determine the importance of the choice of the atmospheric model, 
we must clarify how much uncertainty arises in the inference
of the magnetic field vector when the chosen atmospheric model is different.

It is important to emphasize that
the upper solar chromosphere and transition region are highly
inhomogeneous and dynamic plasmas.
Such inhomogeneity and dynamics causes larger $Q/I$ amplitudes and
non-zero $U/I$ signals, along which their spatial
and temporal variations \citep{Stepan2012,Stepan2014}.
Thus, we must consider also other strategies for interpreting the CLASP observations, 
such as detailed forward modeling of the observed scattering polarization signals
using increasingly realistic 3D models of the solar
chromosphere, taking into account the limited spatial and temporal resolution 
of the CLASP observations. 

In order to monitor the local non-uniformity of the Lyman-$\alpha$ radiation field,
the intensity images from the CLASP slit-jaw and IRIS observations will be useful.
Furthermore, the intensity and the linear polarization profiles in the line wings, which
are insensitive to the magnetic field but very sensitive
to the temperature structure, may also help us to constrain the temperature
structure of the solar atmosphere. All these steps will facilitate the interpretation
of the line-core polarization signals of Lyman-$\alpha$ that CLASP aims at observing. 
In this way, we expect that the CLASP experiment will lead to the first significant advancement in  the investigation of the magnetism of the upper solar chromosphere and the transition region via the Hanle effect in the UV spectral region.

\acknowledgements
This work was done during R.I.'s visit to IAC, which was
supported by the Grant-in-Aid for
``Institutional Program for Young Researcher Overseas Visits''
from the Japan Society for the Promotion of Science (JSPS).
R.I. was also supported by JSPS KAKENHI Grant number 25887051.
J.\v{S}. recognizes support from the Grant Agency of the
Czech Republic through the grant P209/12/P741 and the project
RVO:67985815.
Financial support by the Spanish Ministry of Economy and Competitiveness 
through projects AYA2010--18029 (Solar Magnetism and Astrophysical Spectropolarimetry) and 
Consolider-Ingenio CSD2009-00038 (Molecular Astrophysics: The Herschel and Alma Era)
are gratefully acknowledged. AAR also acknowledges financial support through the Ram\'on y Cajal fellowships.
The CLASP experiment has been finantially supported by a 
Grant-in-Aid for Scientific Research (S).

\end{document}